\documentstyle[preprint,epsfig,aps]{revtex} 

\def\beq{\begin{equation}}
\def\eeq{\end{equation}}
\def\G{{\rm G}}
\def\GN{{G}}
\def\L{\Lambda}

\draft
\tighten
\input epsf

\begin{document}
\draft

\title{Gravitation of the Casimir Effect\\ and the Cosmological Non-Constant}
 

\author{R.~R.~Caldwell}
\address{Department of Physics \& Astronomy \\
Dartmouth College, 6127 Wilder Laboratory\\
Hanover, New Hampshire 03755 USA}

\date{\today}

\maketitle


\begin{abstract}

Whereas the total energy in zero-point fluctuations of the particle physics
vacuum gives rise to the cosmological constant problem, differences in the
vacuum give rise to real physical phenomena, such as the Casimir effect. Hence
we consider the zero-point energy bound between two parallel conducting plates
--- proxy for a solid slice of cosmological constant --- as a convenient
laboratory in which to investigate the gravitation and inertia of vacuum
energy.  We calculate the Casimir effect in a weak gravitational field,
obtaining corrections to the vacuum stress-energy and attractive force  on the
plates due to the curvature of spacetime. These results suggest that if the
cosmological constant is due to zero-point energy then it is  susceptible to
fluctuations induced by gravitational sources.

\end{abstract}

\vspace*{0.2cm}


The Casimir effect \cite{Casimir:1948a,Casimir:1948b} is one of the most
remarkable phenomena in the strange and spooky world of quantum mechanics. At
the most basic level, the attractive force acting on two parallel conducting
plates is a physical manifestation of zero-point energy. Idealized, the plates
are cold and uncharged in vacuo, so there are no real particles or fields
present, but a virtual quantum electrodynamic field creates a pressure
difference across the plates, resulting in this uniquely quantum force.

Our attraction to the Casimir effect follows from its potential  relationship
to the cosmological constant ($\L$) \cite{Weinberg:1988,Carroll:2000} ---
perhaps a cosmological manifestation of zero-point energy. In Zeldovich's
landmark 1968 paper \cite{Zeldovich:1968} he declared that once Einstein has
introduced $\L$, ``the genie has been let out of the bottle and it is no longer
easy to force it back in.'' In a prescient statement, he observed that ``a new
field of activity arises, namely the determination of $\L$'': What is known
about it? What are the limits on it? What experiments can probe $\L$? Numerous
experiments are currently focusing on these very questions!
\cite{DarkEnergyExps} Furthermore, Zeldovich offered the provocative suggestion
that the cosmological constant may be due to the energy density and pressure of
the particle physics vacuum. In support of this speculation, he used the energy
density and pressure for the vacuum of scalar particles of mass $m$
\begin{equation}
\rho = {\hbar c \over 2 (2 \pi)^3} \int_0^\infty \, d^3 k\,
\sqrt{k^2 + (m c/\hbar)^2} , \qquad 
p = {\hbar c \over 6 (2 \pi)^3} \int_0^\infty \, d^3 k\,
{k^2 \over \sqrt{k^2 + (m c/\hbar)^2}}  
\end{equation}
and showed that the regularized stress-energy has the form of Einstein's
constant, whereby $p_{\L}=-\rho_{\L}$. In this paper we take our lead from
Zeldovich, and investigate how the virtual quanta of the vacuum of a scalar
field mixes with gravity.

Where can we find a handful of vacuum energy to analyze in the laboratory? The
reality of zero-point fluctuation forces is firmly established, considering the
dramatic progress which has been made in the measurement of the Casimir effect
in recent years. In fact, the Casimir effect for parallel plates has just
recently been measured, by Bressi and collaborators \cite{Bressi:2002}. (For an
extensive review, see Ref. \cite{Bordag:2001}.) Hence, we will use this
grounded phenomenon as the basis of our investigation, and conduct
``theoretical experiments'' to determine the response of the Casimir effect to
a weak gravitational field. 


We may choose from a variety of methods for calculating the Casimir energy of
parallel plates in Minkowski spacetime, to use as a starting point. We follow
the method described by Milton in Ref. \cite{Milton:2001}, wherein the
stress-energy is obtained from the vacuum expectation values of the
electromagnetic field. In our calculation, however, we will substitute two
free, minimally coupled scalar fields for the two polarization states of the
electromagnetic field.  One scalar field  satisfies Dirichlet boundary
conditions at the surface of the plates, whereas  the other  satisfies Neumann
boundary  conditions  --- in this way the requirements $\hat n \times \vec E=0$
and $\hat n \cdot \vec B = 0$ for the electric and magnetic field on each plate
are met, where $\hat n$ is a unit normal.   For the parallel plate geometry, we
expect our scalar field results to apply to the electromagnetic field. The
central object of our calculation then is the scalar Feynman Green's function,
satisfying $\Box_x \G(x,x') = - \delta(x,x')$, which relates to the
time-ordered scalar field vacuum expectation value $\langle \phi(x)
\phi(x')\rangle = -i \hbar \G(x,x')$.

In the presence of a gravitational field, the Green's function must satisfy
$\Box_x \G(x,x') = -\delta(x,x')/\sqrt{-g}$. Rather than attempt an exact
solution for a given spacetime, we seek an approximate solution which
incorporates the leading-order effects of gravity. In this case, we use the
line-element 
\beq
ds^2 = -(1 - \epsilon \, {2 \GN M \over r}) dt^2 
+ (1 + \epsilon \, {2 \GN M \over r}) d\vec x^2
\label{metric}
\eeq
to describe the weak gravitational field in the vicinity of  a mass $M$, as a
perturbation to Minkowski spacetime. We put the plates at the origin of the
coordinate system, and place the mass at a distance $R$, such that $\GN M/R \ll
1$. It is not necessary to assume that the distance between the plates, $a$, is
small compared to $R$, but we do require that the mass lies outside the plates.
Next, we expand all the terms in the Green's function equation in powers of
$\epsilon$, a dimensionless parameter which measures the strength of the
gravitational field, including the Green's function: $\G(x,x') = \G^{(0)}(x,x')
+ \epsilon \G^{(1)}(x,x') + {\cal O}(\epsilon^2)$.  The superscript indicates
the perturbation order. Matching powers of $\epsilon$, the resultant equations
are
\begin{eqnarray}
&& \Box_x^{(0)} \G^{(0)}(x,x') = -\delta(x,x'), \cr
&& \Box_x^{(0)} \G^{(1)}(x,x') -
{2 \GN M\over r} ( \partial_t^2  + \nabla_x^2 )
\G^{(0)}(x,x') =  {2 \GN M \over r}\delta(x,x').
\label{eqnG0}
\end{eqnarray}
We can use the properties of $\G^{(0)}$ to obtain a solution for $\G^{(1)}$:
\beq
\G^{(1)}(x,x'') = -\int d^4 x' \, \G^{(0)}(x,x') \, {4 \GN M \over r'}
{ \partial_{t'}^2} \, \G^{(0)}(x',x''),
\label{eqnG1}
\eeq
where $\G^{(0)}$ is the appropriate Green's function  for the unperturbed case,
reflecting the boundary conditions enforced at the conducting plates.

The orientation of the conducting plates in the gravitational field is shown in
Figure 1. The distance $r'$ appearing in $\G^{(1)}$ is the distance between a
point $\vec x'$ and the mass at $(R\sin\theta,\,0,\,-R\cos\theta)$.  Since the
integral in equation (\ref{eqnG1}) is dominated by the contribution between the
plates, any orientation dependence enters at order ${\cal O}(\GN M/R^2)$,
consistent with our use of the isotropic metric (\ref{metric}), the weak-field
limit of Schwarzschild. (This answers the speculation \cite{Karim:2000} on an
order ${\cal O}(\GN M/R)$ orientation-dependence correction to the Casimir 
effect in the negative.)

Continuing, the Green's functions are 
\begin{eqnarray}
\G^{(0)}(x,x') &=& \int {d \omega \over 2 \pi}\,{d^2 k \over (2 \pi)^2} \,
{\rm e}^{-i \omega (t-t')+ i \vec k \cdot (\vec x - \vec x')_\perp} 
\, g(z,z')_{D,N} \cr
g(z,z')_D &=&\cases{ 
{\sin\lambda (\frac{a}{2}+z_{<}) \sin\lambda(\frac{a}{2}-z_{>}) \over \lambda \sin\lambda a}
& $-\frac{a}{2} < z, z' < \frac{a}{2}$ \cr
{1 \over \lambda} \sin\lambda(z_< - \frac{a}{2}) {\rm e}^{i \lambda (z_> - \frac{a}{2})}
& $\frac{a}{2} < z, z'$} \cr\cr
g(z,z')_N &=&\cases{ 
-{\cos\lambda (\frac{a}{2}+z_{<}) \cos\lambda(\frac{a}{2}-z_{>}) \over \lambda \sin\lambda a}
& $-\frac{a}{2} < z, z' < \frac{a}{2}$ \cr
{i \over \lambda} \cos\lambda(z_< - \frac{a}{2}) {\rm e}^{i \lambda (z_> - \frac{a}{2})}
& $\frac{a}{2} < z, z'$}
\end{eqnarray}
where the subscript $D,\,N$ indicates Dirichlet or Neumann, and  $\lambda^2 =
w^2 - k^2$. In the absence of the plates, $g(z,z') = i {\rm e}^{i \lambda(z_> -
z_<)}/2 \lambda$ for all $z,\, z'$.  Of course, these are the same reduced
Green's functions used for classical electrodynamics (for example, see Ref.
\cite{Schwinger:1998}), but we will evaluate $\G^{(0)},\,\G^{(1)}$ by a complex
integration which follows the path appropriate for the Feynman Green's
function.  Finally, the stress-energy tensor is 
\beq
\langle T_{\mu\nu} \rangle = \lim_{x' \to x} \,\, \left[
\partial_\mu \partial_{\nu'} - \frac{1}{2} g_{\mu\nu} \, 
\partial_\lambda \partial^{\lambda'} \right] 
\langle \phi(x) \phi(x')\rangle .
\eeq
Thus, we have all the tools necessary to compute the Casimir force and energy.


First, the Casimir force on the plate located at $z=\frac{a}{2}$ is obtained by
evaluating the difference in radiation pressure  on either side, at
$z=\frac{a}{2}^{-}$ and $z=\frac{a}{2}^{+}$. In this way, there is no need to
subtract off an infinite contribution by the background. Taking the
stress-energy tensor and projecting along the unit spacelike normal $\hat
z^\alpha = (0,0,0,1/\sqrt{g_{zz}})$, then the force per unit area $A$ acting on
the plate is finally obtained as
\begin{eqnarray}
{F \over A} = 
\left[ \hat z^\alpha \hat z^\beta \langle T_{\alpha \beta} \rangle 
\right]^{z\to \frac{a}{2}^-}_{z\to \frac{a}{2}^+}  
&=& -{\hbar c \pi^2 \over 240 {a_p}^4} \times  
\left\{  1  +  {\GN M \over R}\left[{2 a_p \over 3 R} \cos\theta 
- {x_p a_p \over R^2} \sin2 \theta     \right. \right. \cr\cr 
&&\left. \left.  
+{2 {a_p}^2(15-\pi^2) \over 9 \pi^2 R^2}  (1 + 3\cos 2\theta) 
+ {\cal O}({\vec x_p \over R})^3 \right] + {\cal O}({\GN M\over R})^2\right\}.
\label{forceresult}
\end{eqnarray}
The subscript $p$ indicates proper length, so $a_p = \int_{-a/2}^{a/2} dz
\sqrt{g_{zz}}$. We have included both the Dirichlet and Neumann fields, which
contribute equally.  For our first result, we find that the attractive force of
the vacuum pressure is modified by the presence of the gravitational field,
with correction terms proportional to the curvature:  components of the Riemann
curvature are proportional to $\GN M/R^3$, {\it e.g.}   $R_{tztz} =
-\frac{1}{2}(1 + 3 \cos 2\theta)\GN M / R^3 $. For $\theta=0$ the force is
enhanced, whereas for $\theta = \pi/2$ the magnitude of the force is weakened. 
Hence, the pressure of vacuum energy responds to variations in the neighboring
gravitational field.


Second, the Casimir energy is obtained by projecting the stress-energy tensor
along the unit timelike vector $\hat t^\alpha = (1/\sqrt{|g_{tt}|},0,0,0)$ and
integrating over the volume between the plates.  In this case, we must subtract
the energy density in the absence of the plates, as well as remove  
unphysical, divergent contact terms due to the presence of the plates. (See
Refs. \cite{Deutsch:1979,Kennedy:1980}.) It is interesting to note that the 
Dirichlet and Neumann contact terms cancel to ${\cal O}(\GN M/R)$. This
cancellation does not carry over to higher order terms. However, if we consider
the alternative stress-energy tensor ${\widetilde T}_{\mu\nu} = -\phi
\nabla_\mu\nabla_\nu \phi + \frac{1}{2} g_{\mu\nu} \phi \Box \phi$,  which
differs from the standard form by surface terms, then we can use a linear
combination of $T,\, {\widetilde T}$ to eliminate the unphysical, diverging
contact terms. In this way, the stress-energy tensor is found to be
\begin{eqnarray}
\langle {T^\alpha}_\beta \rangle &=& {\hbar c  \pi^2 \over 720 {a_p}^4}
\left\{ {\bf M}_0  + {\GN M \over R} \times \right.  \cr\cr
&& \left[ {\bf M}_1 \left( {4 z_p \over 5 R} \cos\theta   
-     {6 x_p z_p \over 5 R^2}\sin 2\theta
+ {a_p^2 - 3(z_p + \frac{a_p}{2})^2 \over 15 R^2}
{(1 + 3 \cos 2 \theta)} \right) \right. \cr\cr
&& \left.\left.
+ {\bf M}_2 {a_p^2 \over R^2}{1 \over \pi^2}  
+ {\bf M}_3 {a_p^2 \over R^2}{\cos 2\theta \over \pi^2}  
+ {\bf \Pi}{3 \over 10} {(a_p^2 - 4 z_p^2) \over R^2}\sin 2\theta
\right]\right\}
\label{stressenergyresult}
\end{eqnarray}
where the diagonal matrices are
${\bf M}_0 = {\rm diag}(-1,1,1,-3)$, 
${\bf M}_1 = {\rm diag}(1,3,3,-5)$,
${\bf M}_2 = {\rm diag}(-3,8,-13,-10)$,
${\bf M}_3 = {\rm diag}(-9,-18,3,-30)$,
and the anisotropic term is ${\bf \Pi}|^{x}_{z} = {\bf \Pi}|^{z}_{x} = 1$,
${\bf \Pi}=0$ otherwise. The ${\bf M}_0$ term above is consistent with  the
results for the QED stress-energy tensor   derived by Brown \& Maclay
\cite{Brown:1969} using the symmetries of the electromagnetic vacuum state in
the presence of the conducting plates,  to the extent that we use the proper 
plate separation, $a_p$, in the absence of curvature or finite-size effects 
(also see \cite{Calloni:2001}). The above stress-energy tensor
(\ref{stressenergyresult}) is covariantly conserved, ${T^\mu}_{\nu;\mu} = 0$.
However, ${\bf M}_{2,3}$ are not traceless, and the stress-energy acquires
a trace
\begin{equation}
T = g^{\mu\nu} T_{\mu\nu} = -{ \hbar c \over
60 {a_p}^4} {\GN M \over R}\left[ {a_p^2 \over R^2}(1 + 3 \cos 2\theta) 
+ {\cal O}({\vec x_p \over R})^3 \right] + {\cal O}({\GN M\over R})^2.
\label{traceresult}
\end{equation}
Whereas the symmetry of the plates and the boundary conditions keep the
tree-level stress-energy tensor traceless for our minimally-coupled field, the
presence of the gravitational field breaks the planar-symmetry. The
curvature-induced distortions of the fields now serve as a source for further
gravitation. 
 

Third, the field energy stored in the Casimir plates is
\begin{eqnarray}
{\cal E} &=& \int d^3\Sigma \,\, \hat t^\alpha  \hat t^\beta
\langle T_{\alpha \beta} \rangle  
= \int (\sqrt{-g} / \sqrt{-g_{tt}}) \, d^3 x \, \hat t^\alpha  \hat t^\beta \, 
\langle T_{\alpha \beta} \rangle \cr\cr
&=& -{\hbar c  \pi^2 \over 720 {a_p}^4} (A_p a_p)
\left\{ 1 + {\GN M \over R}
\left[{{a_p}^2 \over \pi^2 R^2}(1 + 3\cos 2\theta) 
+ {\cal O}({\vec x_p \over R})^3 \right]
+ {\cal O}({\GN M\over R})^2  \right\}.
\label{energyresult}
\end{eqnarray} 
Here, $d^3\Sigma$ is the 3-volume element for an observer with 4-velocity $\hat
t^\alpha$ at fixed time, and the proper area $A_p$ is a circular disk in the
x-y plane centered at the origin. We have included both the Dirichlet and
Neumann fields. As with the force, the curvature enhances the  energy stored in
the plates for $\theta = 0$, and diminishes the energy for $\theta = \pi/2$. 
Hence, the vacuum energy  responds to variations in the neighboring
gravitational field.
  

The weak gravitational field exerts a force on the Casimir plates, as well as
the vacuum energy and pressure bound to the plates. Neglecting tidal forces,
the plates themselves would fall as a rigid body along a geodesic towards the
mass $M$. Focusing on the fields alone, as a theoretical experiment we may
consider the force necessary to support the energy-momentum of the   fields in
the vicinity of the mass which we can interpret as the ``weight of the vacuum''
\begin{equation}
w_{\rm C} = \hat r^\alpha f_\alpha =\hat r^\alpha \hat t^\mu 
(\int  d^3 \Sigma \, \hat t^\beta \, \langle T_{\alpha\beta} \rangle
)_{;\mu}
= {\cal E} {\GN M \over R^2}
\left[1 - {\GN M \over R} + {\cal O}({\GN M\over R})^2 \right]. 
\label{weighteqn}
\end{equation}
Here, $f^\alpha$ is the 4-force which we have contracted  with a unit 4-vector
$\hat r$, pointing from the origin to the mass, $M$. (Because the leading term
in the covariant derivative is already of order $\GN M/R^2$, we have included
one more term in the weak field expansion of the metric, namely $g_{tt} = -1 +
2 (\GN M/r) - 2 (\GN M/r)^2$, consistent with the Schwarzschild metric in
isotropic coordinates. Hence, the expression above is the standard
parametrized-post-Newtonian result \cite{Will:1993}.) For comparison, a point
particle of mass $m$ has $f^\alpha = u^\mu {p^\alpha}_{;\mu}$ with $p^\alpha =
m u^\alpha$, so the weight is identical to (\ref{weighteqn}) with the
substitution ${\cal E} \to m$.  Of course, the acceleration is the same, but
the weight is influenced by the orientation of the plates relative to the mass,
due to the $\theta-$dependence in equation (\ref{energyresult}). 

The vacuum energy would seem to be buoyant, since $w_{\rm C}<0$. If we regard
the negative Casimir energy as a sort of particle with negative gravitational
mass, we would playfully expect it to accelerate towards the mass $M$, which in
turn would be gravitationally repelled by the Casimir particle. One particle
chases the other. (See problem 13.20 of Ref. \cite{Lightman:1975}.) Although
the conducting plates contribute a positive mass $m$ to the system, we
might imagine very light plates for which $m + {\cal E} <0$, to
maintain our negative mass object. Nevertheless, the strain required to oppose
the Casimir force and keep the plates at a constant separation contributes an
energy density at least as large as the pressure (assuming the mechanism
holding the plates fixed is composed of a material which satisfies the weak
energy condition), so that the net mass-energy is positive. The Casimir energy
is a binding energy, after all. This recalls Misner and Putnam's discussion of
active gravitational mass \cite{Misner:1959}: although the energy density and
pressure of the fields serve as a source of gravitation, via $\Box h_{\mu\nu} =
-16 \pi G (T_{\mu\nu} - \frac{1}{2}g_{\mu\nu} T)$, if the pressure is contained
by non-gravitational means, then the associated tension must offset the
pressure, so that for the isolated system the active gravitational mass and
inertial mass are the same. 

The distinction between gravitation and inertia brings us back to the first
cosmological models in Einstein's nascent Theory of General Relativity. On the
one hand we have de~Sitter's world-model, wherein the cosmological spacetime is
filled by a ``world-matter'' ($\L$) responsible for inertia, whereas the
sources of gravitation,  stars and nebulae like massive particles moving in the
fixed background, are to be considered as gravitational perturbations  
\cite{deSitter:1917}. This may seem to be an apt description of a Universe
dominated by an idealized, immutable, yet unphysical cosmological constant. On
the other hand, Einstein had supposed that the sun and stars may be condensed
``world-matter,'' and that the world-matter itself may very well be
inhomogeneous on small scales, approaching a constant density on sufficiently
large scales \cite{Einstein:1917}. We can readily apply Einstein's description
to our results for the Casimir vacuum field energy, as a proxy for a
physically-motivated cosmological constant.

We still don't know how to calculate $\L$ from first principles, but if there
does exist a finite $\L$ arising from zero-point energy, then we should not
expect it to be a true and immutable constant. Although the presence of a
boundary in the form of conducting plates is essential to the Casimir effect,
the gravitational distortion is not simply a consequence of a distorted
boundary. True, the boundary planes are necessary in order for us to compute
something finite. But the expectation $\langle \phi(x) \phi(x')\rangle$ in the
absence of any boundaries is itself influenced by gravitational sources,
following equations (\ref{eqnG0},\ref{eqnG1}). If we could compute other finite
quantities involving vacuum energy,   they too would experience a distortion
due to the presence of a gravitational field. The lesson from our calculations
is that the vacuum is not rigid, but instead is susceptible to fluctuations
driven by gravity.

The magnitude of the effects we have considered for the system of Casimir
plates are too small to be detected in the laboratory, although strong
curvature as found near tiny black holes with Schwarzschild radius $r \sim a$  
would produce a significant distortion, {\it e.g.} an Earth-mass black hole for
$a\sim 1~$cm. (See also \cite{Parker:1980}.) In the cosmological scenario,
strong curvature can be expected to produce a distortion of $\L$, although the
relevant length scales remain obscure. If the dimensionless corrections go like
$\GN M a^2/R^3$, and  $R$ gives the distance to a nearby mass, what is $a$? A
length scale associated with a cosmological boundary or horizon? If so, then
the gravitational corrections may be substantial. On the other hand, a length
scale $a \sim (\hbar c/\rho_c)^{1/4}$ where $\rho_c$ is  the critical density
of the Universe would give negligible corrections.   Needless to say,
detection of a large-scale variation in the cosmological constant would have
tremendous impact on our understanding of the nature of the cosmic vacuum. 


An outstanding challenge to physics is whether vacuum energy or virtual
particles obey the equivalence principle, fall along geodesics, and obey
mechanical laws in analogy to classical material. Some of these questions may
be within reach of experiment in the near future. (See Refs.
\cite{Jaekel:1993,Viola:1996,Jaekel:1997,Papini:2001}.) Ultimately, the answers
will help us understand the physics of the quantum world. We may learn whether
vacuum energy contributes to inertia and gravitation, and possibly what is the
role of vacuum energy as a cosmological (non-)constant.

 
\acknowledgements 
We thank Miles Blencowe, Larry Ford, and Marcelo Gleiser for useful
conversations. This work was supported in part by Research Corporation grant
RI-0887 and NSF grant PHY-0099543. A detailed exposition of the calculations
contained in this paper is forthcoming.


\begin{figure}
\begin{center}
\epsfxsize=2.0 in \epsfbox{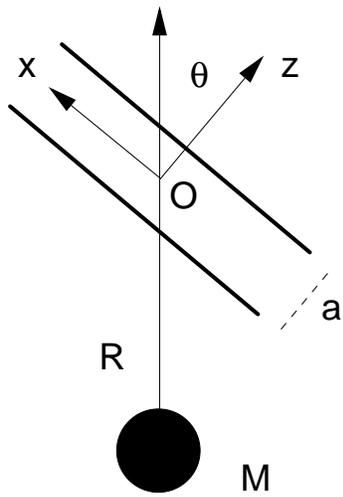}
\end{center}
\caption{The orientation of the parallel conducting plates is shown. The origin
of the coordinate system is marked by the letter O, and the mass M is located
at $(-R\sin\theta,\,0,\,-R\cos\theta)$.
\label{figure1}}
\end{figure}


\end{document}